# A Chaotic Cipher Mmohocc and Its Randomness Evaluation

**Xiaowen Zhang[a], Ke Tang[a], Li Shu[b]**

[a] Dept. of CS, the Graduate Center, CUNY
[b] College of Computer Science, Sichuan University
xzhang2(at)gc.cuny.edu

## Abstract

After a brief introduction to a new chaotic stream cipher Mmohocc which utilizes the fundamental chaos characteristics of mixing, unpredictability, and sensitivity to initial conditions, we conducted the randomness statistical tests against the keystreams generated by the cipher. Two batteries of most stringent randomness tests, namely the NIST Suite and the Diehard Suite, were performed. The results showed that the keystreams have successfully passed all the statistical tests. We conclude that Mmohocc can generate high-quality pseudorandom numbers from a statistical point of view.

## 1 Introduction

Claude E. Shannon [Shannon 1949] indicated that stretch-and-fold mechanism of mixing transformation is an important component of a cipher and it is also a way to confuse and diffuse messages. The chaotic dynamical systems, which exhibit unpredictable, mixing, and extremely sensitive to initial conditions, have shown strong connection with cryptography since its inception. It has attracted a broad interest among cryptographic researchers to investigate the cryptographic applications for chaotic properties [Baptista 1998, Alvarez 1999, Jakimoski 2001, Fridrich 1998, Kocarev 1998]. However the properties of normal chaos systems are based on continuous systems and they are long-term behaviors. Due to limited resource and computational power in digital systems those properties will be severely degraded or even disappeared. In order to use chaotic properties in cryptosystems, we have devised a technique which accelerated chaotic behaviors more rapidly than those in a normal chaos system.

The Mmohocc [Zhang 2006] (an acronym, pronounced "mow-hock"), *m*ulti-*m*ap *o*rbit *ho*pping *c*haotic *c*ipher, is still under development. Its primary goal is to exploit basic properties of chaos systems to design a cryptographically strong, fast, and feasible cipher. By using multiple maps, i.e., multiple chaos systems, the Mmohocc cipher generates an extremely long chaotic sequence inspired by Vernam's one-time pad. By hopping among multiple orbits generated from multiple maps the cipher obtains its confusion and diffusion properties in a much faster way, which is a basic asset for any good cryptosystem.

A chaotic map *F* is a map / function, usually a non-linear discrete dynamical iteration equation, which exhibits some sort of chaotic behavior. A chaotic orbit is the trajectory that a chaotic map iterated. Given $x_0 \in R$ and a chaotic map *F*, we define the orbit of $x_0$ under $F$ to be the sequence of points $x_0$, $x_1 = F(x_0)$, $x_2 = F^2(x_0)$, $x_n = F^n(x_0)$, …. The point $x_0$ is called the seed of the orbit [Devaney 1992].

How do we evaluate keystreams generated by the Mmohocc? There are batteries [NIST Computer Security Division 2001] of statistical tests available to analyze cryptographic random number generators (RNGs) and pseudorandom number generators (PRNGs): the pLab, the Crypt-X, the DIEHARD, the NIST, the ENT, the RIPE, and others. Among them the NIST and the DIEHARD are considered the most stringent randomness tests. To assess the Mmohocc cipher we conducted the two aforementioned batteries of tests and the results showed that our cipher has successfully passed all of them.

## 2 The Mmohocc – A Chaotic Stream Cipher

The Mmohocc is a software based stream cipher. It "leaps" among substantially many chaotic orbits to "pick up" its random sequence points. The base ground for random sequences is spread among lots of chaotic orbits, which are generated from multiple chaotic maps. The notion of orbit-hopping is also inspired from the frequency-hopping (FH) communication. It is one of the modulation methods in spread spectrum communications, in which radio signals, following a hopping pattern, switch a carrier among many frequencies. A hopping pattern is a predefined pseudorandom sequence known to both transmitter and receiver [Wikipedia].

Fig. 1 shows the constructing blocks of the Mmohocc. The Key Scheduling expands a 128-bit (or 256-bit, or 512-bit) key into many subkeys (SKs) for controlling the operations of multiple chaotic maps. Each subkey contains certain number of fields. Among them a field called hpsn (hopping-pattern serial number) is used by the Hopping Mechanism to govern the jumping behavior between the chaotic orbits. In the following subsections we will explain the hopping-pattern and the random number extraction.

### 2.1 Hopping Patterns

The hopping pattern is a predefined pseudorandom sequence of certain number of orbits, it tells how Mmohocc hops among orbits. In the current version the hpsn field of a subkey takes 8 bits long, therefore it can be changed between 0 and 255. Each hpsn corresponds to one of available hopping patterns, which are stored in a lookup table. For instance, if a chaotic map has 11 orbits and its hopping-pattern is {7 3 9 1 6 10 4 5 2 11 8}, then we know that when it comes to this map Mmohocc extracts random numbers on orbit 7 first and then orbit 3, orbit 9, … orbit 11 in this order.

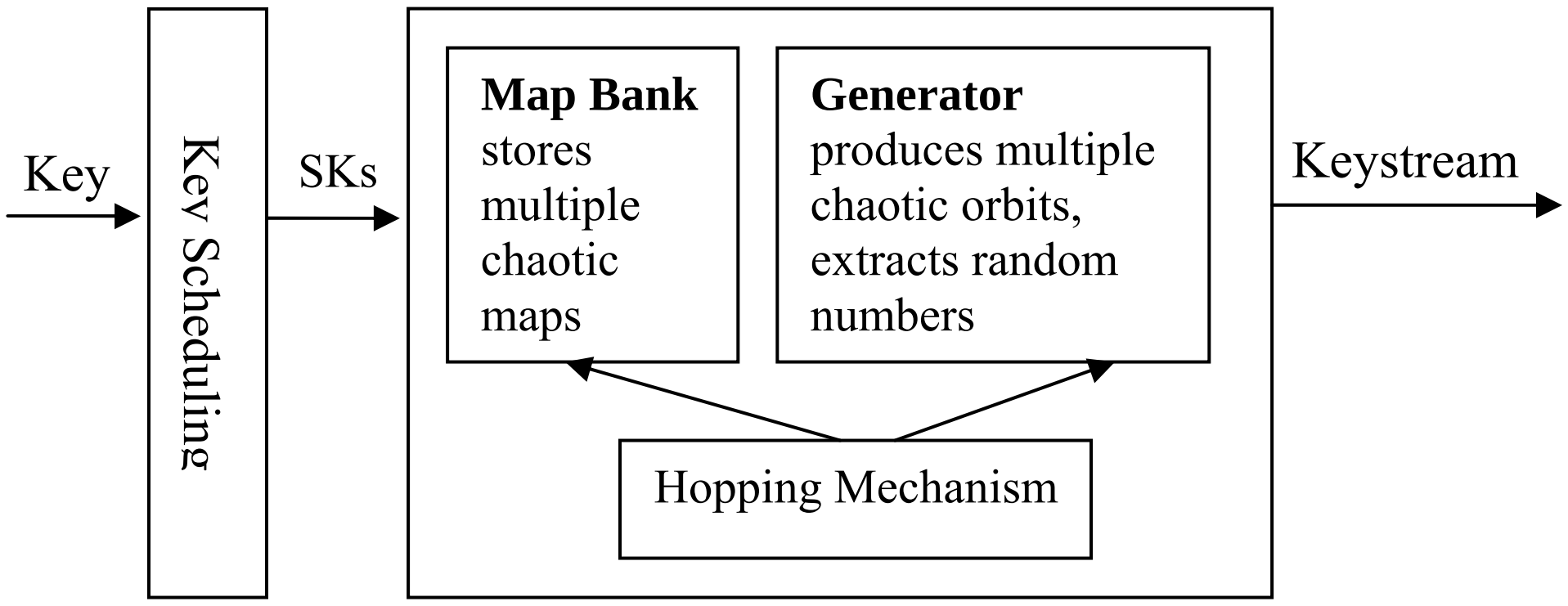

Fig. 1 Block Diagram of the Mmohocc

### 2.2 Random Number Extraction

A chaotic orbit point $x_n$ is a real number (in the case of logistic map, $x_n < 1.0$) and is represented in a data type of double in computer. We extract two smaller integers as random numbers from the point $x_n$ in Fig. 2 method. By doing so, we further muddle the bits and add additional randomness to the keystream.

- Step 1: Convert double type $x_n$ into a 32-bit integer by multiplying $2^{52}$.
- Step 2: Split the above integer into four 8-bit integers *a*, *b*, *c*, *d*. Here *a* and *d* are the leftmost and rightmost 8-bit, *b* and *c* are in the middle. Then the output two smaller pseudorandom integers are: $u = a \oplus c$ and $v = b \oplus d$.

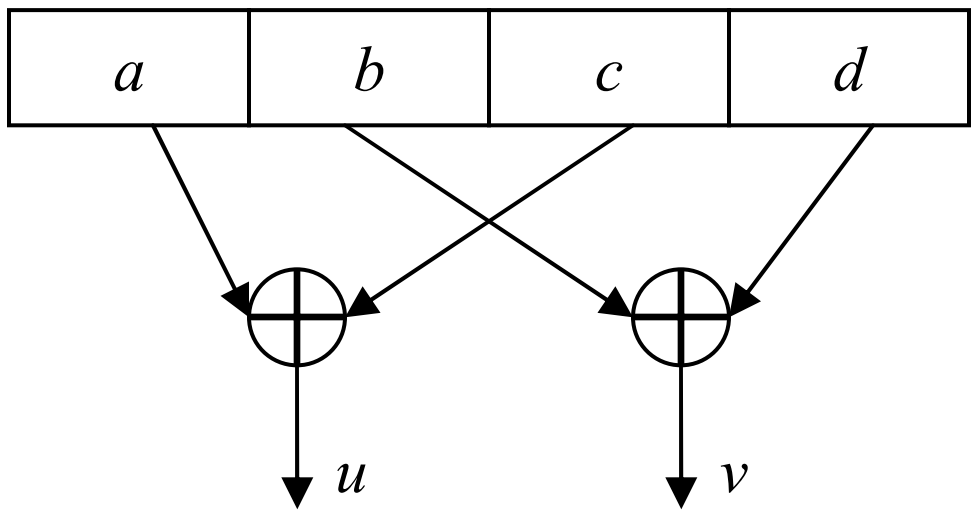

Fig. 2 Random Number Extraction

## 3 Statistical Tests

The randomness of a bit sequence is characterized and described in terms of probability. The NIST Suite and the Diehard Suite are highly regarded and free of charge among the list of batteries of tests. They both include dozens of independent and computationally intensive statistical tests. Most of these tests return a test statistic and its corresponding probability value (*p*-value) [Soto]. The *p*-value [Widipedia] is the probability of obtaining a test statistic as "impressive" as the one observed if the sequence is random, so that the statistic was the result of chance alone. In other words, the *p*-value summarizes the strength of the evidence against the perfect randomness hypothesis. Small values ($p$-values < 0.05 or $p$-values < 0.01) are interpreted as evidence that a sequence is unlikely to be random. Here 0.05 and 0.01 are significance level, usually denoted as α.

The $p$-values are obtained by $p = F(z)$, where $F$ is a special function such as the complementary error function *erfc*, the incomplete gamma function *gamminc*, the standard normal (cumulative distribution) function *normcdf*, or the gamma function *gamma*.

### 3.1 The NIST Statistical Test Suite

The NIST Suite [Rukhin 2001, NIST Suite] provides a battery of 16 statistical tests. They assess the presence of a pattern which, if detected, would indicate that the sequence is non-random. In each test a $p$-value is calculated. The significance level α for all tests in NIST Suite is set to 1%. A $p$-value of zero indicates that the sequence appears to be completely non-random. A $p$-value less than α would mean that the sequence is non-random with a confidence of 99%. If a $p$-value is greater than α, we accept the sequence as random with a confidence of 99%.

Table 1 shows a result obtained on a series of data files of 1,192,755,216 (over 1 billion) bits each. The files were generated from the Mmohocc by using different keys. The Suite took one of files as input data, and executed all 16 tests. Set 1000 as the number of bit sequences, each containing 1 million random bits. The mean and variance of the $p$-values are displayed in the table.

Table 1 Mean and Variance of $p$-values for A Large Bit Sequence

| TSN | Test Name | Mean of p-value | Variance | Conclusion |
|---|---|---|---|---|
| 1 | Approximate Entropy | 0.5146 | 0.0857 | Success |
| 2 | Block Frequency | 0.5040 | 0.0844 | Success |
| 3 | Cumulative Sums (Forward) | 0.4777 | 0.0799 | Success |
| | Cumulative Sums (Reverse) | 0.4824 | 0.0815 | Success |
| 4 | Fast Fourier Transform (Spectral) | 0.4799 | 0.0800 | Success |
| 5 | Frequency (Mono-bit) | 0.4894 | 0.0827 | Success |
| 6 | Lempel-Ziv Compression | 0.5024 | 0.0823 | Success |
| 7 | Linear Complexity | 0.5024 | 0.0823 | Success |
| 8 | Longest Runs of Ones | 0.5121 | 0.0849 | Success |
| 9 | Maurer's Universal Statistical | 0.5000 | 0.0889 | Success |
| 10 | Non-Overlapping Template Matching | 0.4997 | 0.0833 | Success |
| 11 | Overlapping Template Matching | 0.4970 | 0.0830 | Success |
| 12 | Random Excursions | 0.5087 | 0.0835 | Success |
| 13 | Random Excursions Variant | 0.5103 | 0.0813 | Success |
| 14 | Rank | 0.4963 | 0.0841 | Success |
| 15 | Runs | 0.4980 | 0.0841 | Success |
| 16 | Serial | 0.5024 | 0.0847 | Success |

(TSN: Test Serial Number)

For further interpreting empirical results NIST Suite has also adopted the following two approaches.

(1) The Examination of the Proportion of Passing Sequences
In the final analysis report file generated by the suite, a value called the proportion was listed for each test. The proportion is the number of sequences having a *p*-value greater than the significance level α, divided by the total number of bit sequences tested. That is the percentage of passed tests.

NIST SP800-22 specifies a range of acceptable proportions. The range is determined by using the confidence interval defined as, $\hat{p} \pm 3\sqrt{\hat{p}(1-\hat{p})/m}$, where $\hat{p} = 1-\alpha$, and $m$ is the sample size, which tells how many bit sequences were tested.

In our case, 1000 bit sequences ($m$ = 1000) and 1 million bits per sequence were used. By the formula above, the range of acceptable proportion is from 0.9805 to 0.9994, inclusively. Fig. 3 shows the proportion for each of the 16 tests. Since the proportion for each test is within the range, so we are confident to accept the sequence as random bit sequence.

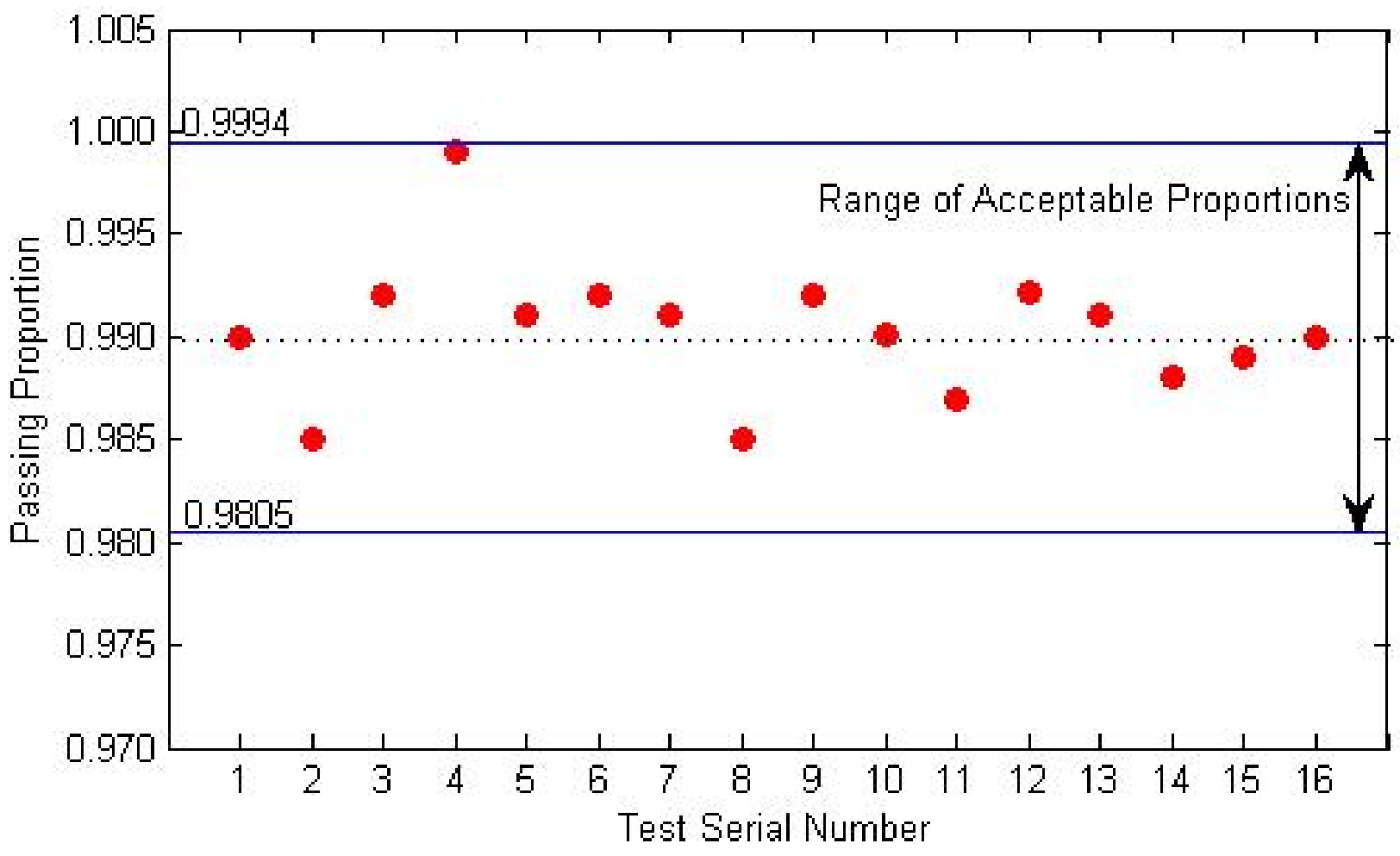

Fig. 3 Passing Proportions of 16 NIST Tests

(2) The Examination of the Uniformity of *p*-values
For visually examining the uniformity of *p*-values, we can make a histogram for each test. The horizontal coordinate (the probabilities) interval between 0 and 1 is divided into 10 sub-intervals evenly, and the number of *p*-values that drops within each sub-interval is displayed. As an example, Fig. 4 shows the histograms of *p*-values for longest runs and rank tests. We can see that the *p*-values are uniformly distributed.

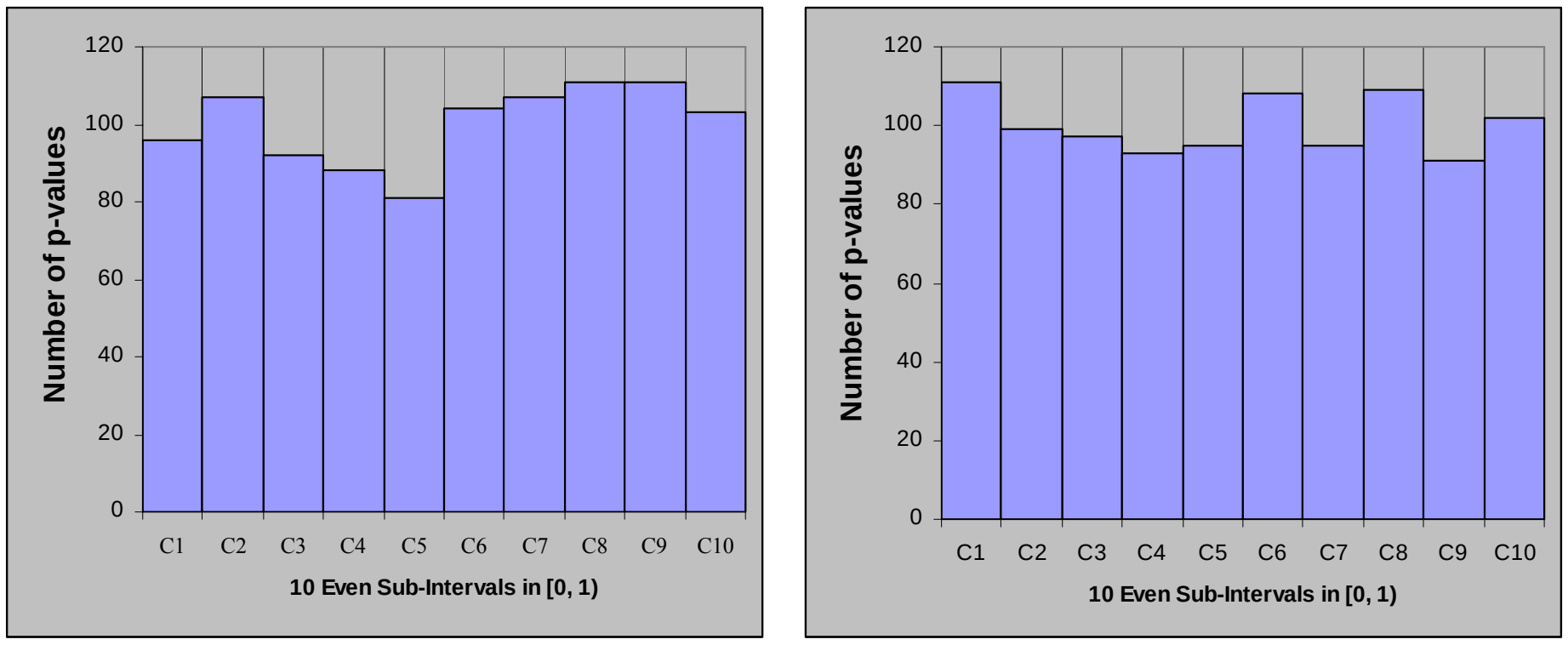

Fig. 4 Histogram of p-values for Longest Runs and Rank

Uniformity may also be examined by computing the following $\chi^2$ value for each test,

$$\chi^2 = \sum_{i=1}^{10} \frac{(Ci - m/10)^2}{m/10}$$

where $Ci$ is the number of $p$-values in sub-interval $[(i\text{-}1)/10, i/10)$ ($i = 1 \sim 10$), and m is the sample size (i.e., the number of bit sequences tested). We calculate a $p$-value of the $p$-values obtained for a statistical test as

$$p_p\text{-value} = gammainc(\chi^2/2, 9/2, \text{'}upper\text{'}) .$$

If $p_p \geq 0.0001$, then the $p$-values can be considered uniformly distributed. Table 2 shows the $p_p$-values of the 16 tests and all $p$-values are uniformly distributed.

Table 2 Uniformity Distribution of $p$-values

| TSN | 1 | 2 | 3 | 4 | 5 | 6 | 7 | 8 |
|---|---|---|---|---|---|---|---|---|
| $p_p$-value | .546 | .432 | .035 | .057 | .197 | .047 | .693 | .392 |
| Conclusion | Pass | Pass | Pass | Pass | Pass | Pass | Pass | Pass |
| TSN | 9 | 10 | 11 | 12 | 13 | 14 | 15 | 16 |
| $p_p$-value | .302 | .473 | .347 | .395 | .380 | .868 | .107 | .508 |
| Conclusion | Pass | Pass | Pass | Pass | Pass | Pass | Pass | Pass |

### 3.2 The Diehard Suite

The Diehard Suite developed by George Marsaglia [Marsaglia 1995] consists of 18 stringent statistical tests. Binary file must be provided – a file of 10 to 11 megabytes, i.e., at least 80 million bits. 100 binary files to be tested of 100 million bits each were generated by Mmohocc with 100 different keys. Table 3 shows a result produced from a single binary file of 100 million bits. A *p*-value larger than 0.01 and smaller than 0.99 means that the sequence is random with a confidence of 99%. All the testing results obtained on the Mmohocc confirmed that a high level of confidence in the randomness of the keystreams has been achieved.

Table 3 *p*-values from the Diehard Statistical Test Suite

| TSN | Test Name | *p*-value | Conclusion |
|---|---|---|---|
| 1 | Birthday Spacing | 0.3213 | Success |
| 2 | Overlapping 5-Permutation | 0.1489 | Success |
| 3 | Binary Rank (31 x 31 Matrices) | 0.8363 | Success |
| 4 | Binary Rank (32 x 32 Matrices) | 0.3227 | Success |
| 5 | Binary Rank (6 x 8 Matrices) | 0.4506 | Success |
| 6 | Bitstream | 0.5638 | Success |
| 7 | Overlapping-Pairs-Sparse-Occupancy | 0.5704 | Success |
| 8 | Overlapping-Quadruples-Sparse-Occupancy | 0.5193 | Success |
| 9 | DNA | 0.4021 | Success |
| 10 | Count-The-1's (on stream of bytes) | 0.2279 | Success |
| 11 | Count0-The-1's (on specific bytes) | 0.5343 | Success |
| 12 | Parking Lot | 0.9362 | Success |
| 13 | Minimum Distance | 0.2327 | Success |
| 14 | 3D Spheres | 0.3366 | Success |
| 15 | Squeeze | 0.4369 | Success |
| 16 | Overlapping Sums | 0.6415 | Success |
| 17 | Runs | 0.5000 | Success |
| 18 | Craps | 0.5579 | Success |

## 4 Conclusion

By rapidly hopping among many orbits/many maps the Mmohocc cipher greatly speeds up the chaotic mixing. This makes the Mmohocc a feasible stream cipher. Although evaluating statistical characteristics of keystreams can never replace security analysis for a stream cipher, in principle it is an inevitable and standard procedure to go through randomness tests. Successfully passing all two batteries of the stringent statistical tests confirms our design, supports that. We conclude the paper with the following summary.

- The Mmohocc used a hopping mechanism to implement a chaotic cipher.
- The keystreams generated by the Mmohocc have passed two batteries of randomness tests with satisfactory results.
- The Mmohocc generated high quality pseudorandom numbers.

- The randomness statistical tests ensured our technique and helped to improve our adjustments for certain parameters/coefficients involved in the chaotic maps, orbit hopping offsets, and most importantly hopping patterns.

## Acknowledgements

The authors would like to thank Dr. Lin Leung for her scrupulous review and Prof. Michael Anshel for his helpful comments and encouragement.